\documentclass[]{interact}

\usepackage{epstopdf,subfiles}
\usepackage[pagewise]{lineno}
\usepackage[caption=false]{subfig}
\usepackage{mathrsfs,amsmath,amsfonts,float,appendix}

\usepackage[numbers,sort&compress]{natbib}
\bibpunct[, ]{[}{]}{,}{n}{,}{,}

\newcommand{\ceil}[1]{\lceil #1 \rceil}
\newcommand{\floor}[1]{\lfloor #1 \rfloor}

\theoremstyle{plain}

\theoremstyle{definition}

\theoremstyle{remark}

\begin{document}
\title{Investigating the Trade-off between Infections and Social Interactions Using a Compact Model of Endemic Infections on Networks}

\author{
{Bunlang Thatchai\textsuperscript{a,$\dagger$}, Christopher E. Overton\textsuperscript{b,c}, and Thomas House\textsuperscript{a}}
\thanks{{}\textsuperscript{$\dagger$}Corresponding author: bunlang.thatchai@manchester.ac.uk}
\vspace{6pt}\affil{\\ \textsuperscript{a}Department of Mathematics, School of Natural Sciences, The University of Manchester, Manchester, UK.\\ \textsuperscript{b}Department of Mathematical Sciences, School of Physical Sciences, University of Liverpool, Liverpool, UK.\\
\textsuperscript{c}Modelling Division, Analysis and Intelligence Assessment, Chief Data Officer Group, UK Health Security Agency, London, UK.}
}

\maketitle
\begin{abstract}
In many epidemiological and ecological contexts, there is a trade-off between infections and interactions. This arises because the links between individuals capable of spreading infections are also often associated with beneficial activities. Here, we consider how the presence of explicit network structure changes the optimal solution of a class of infection-interaction trade-offs. In order to do this, we develop and analyse a low-dimensional dynamical system approximating the network SIS epidemic. We find that network structure in the form of heterogeneous numbers of contacts can have a significant impact on the optimal number of contacts that comes out of a trade-off model.
\end{abstract}
\begin{keywords}
Susceptible-infective-susceptible model;
Pairwise approximation;
Endemic equilibrium; 
Random graph; 
Benefit of individual interaction;
Cost-benefit analysis.

\end{keywords}

\section*{Plain Language Summary}
Humans and other species benefit from interacting with each other, but at the same time such contact can spread infection. This can lead to a trade-off between the
costs and benefits of contact. Here, we develop a parsimonious mathematical model of susceptible-infectious-susceptible (SIS) epidemic dynamics on a social contact network.
We then use this to investigate the impacts of the structure of such contact networks on infection-interaction trade-offs given different models of benefits. We find that in many cases the quantitative and qualitative impacts of contact network structure are important for quantifying these trade-offs.

\section{Introduction}

The COVID-19 pandemic posed the question of trade-offs between disease control and general well-being with particular intensity, seeing both millions of deaths due to the disease itself \cite{WHO:2024}, and unprecedented measures taken by Governments to curtail contacts between people and hence transmission \cite{Hale:2021}. Similar challenges are seen more widely at other times in human history and throughout ecology \cite{Whittles:2016, Herrera:2019}. While COVID-19 caused more relevant data to be available for consideration of infection-interaction trade-offs than was previously possible \cite{Vedhara:2022, Wu:2022, Bolton:2024}, there is also a need for more theoretical understanding of these, and that is the aim of this work.

The role of contact networks has been widely explored in infectious disease epidemiology, due to the key role of contact-based interactions in spreading many infectious diseases \cite{Eames:2002, Hagenaars:2004, Keeling:2005, Kiss:2017, Pastor-Satorras:2001}. The structure of the contact network can have major implications on the spread of the disease, such as altering the epidemic threshold \cite{Ferreira:2012, Kiss:2017} and the final size \cite{Kiss:2017}. A challenge with the incorporation of network structure into epidemic models is the rapid escalation in computational costs of the models. These make the systematic analysis of network structure infeasible for the vast majority of networks, with very few networks having analytical tractability. To improve analytical capability, moment-closure methods, such as pair-approximation, have been developed to approximate the expected dynamics of the model \cite{Eames:2002, Lajmanovich:1976, Sharkey:2011}. These models exhibit endemic equilibrium solutions \cite{Kiss:2017} that have been shown to well-approximate the quasi-stationary distributions (QSD) of the SIS model \cite{Overton:2022}. This is a valuable approximation since for large networks above the epidemic threshold, although the QSD is not a genuine equilibrium in the stochastic epidemic, the time to extinction can be long enough that the QSD describes the typical long-term behaviour of the system \cite{Andersson:2000, Kryscio:1989, Naasell:1996, Wilkinson:2013}.

Within evolutionary dynamics, network structures have also been widely studied as a method for considering how interaction structure affects competition between mutants \cite{Lieberman:2005}. Within these studies, evolutionary game theory can be used to assign benefits and costs for the number of interactions an individual makes, which can have large implications on the evolutionary dynamics, such as amplifying or suppressing the rate of natural selection \cite{Lieberman:2005, Masuda:2009, Ohtsuki:2006, Broom:2010, Hindersin:2015}. 

In this paper, we aim to explore, within a theoretical framework, the trade-off between social connectivity, the associated benefits linked to each interaction, and infectious disease resilience, with the associated costs of increased connectivity increasing the size of the endemic equilibrium. To achieve this, we consider the
susceptible-infectious-susceptible (SIS) epidemic model on a network in which individuals do not
exhibit immunity after recovery, and which can be used in epidemiological contexts for modelling sexually transmitted
infections \cite{Eames:2002}. This model also has the benefit for theoretical work of relative simplicity and existence of a non-zero prevalence endemic state. To analyse the model, we use a pair-approximation moment closure approach to allow efficient analysis of general network structures. A pair-based approach is essential since mean-field equations alone are not sufficient to analyse the influence of network structure on the costs and benefits within the model. To perform this analysis, we develop a new method to derive the endemic equilibrium of the pair-approximation model, enabling efficient calculation of the influence the network structures have on the endemic equilibrium. We trade this off against a family of benefit functions that are sigmoidal in number of contacts. 

\section{Definitions and analytical results}

\subsection{Networks}

A network $G=(\mathcal{V}, \mathcal{L})$ comprises a collection of vertices $v_i\in\mathcal{V}$, $i\in [\mathcal{V}]$ and edges in $\mathcal{L} \subseteq \mathcal{V}\times\mathcal{V}$ where $e_{ij} = (v_i, v_j) \in\mathcal{L}$ denotes an edge connecting the vertex $v_i$ and $v_j$. Vertices and edges can equivalently be called \textit{nodes} and \textit{links} respectively. We consider a graph $G \in \mathcal{G}$ with $N$ nodes, indexed by integers $i, j, \ldots \in \{ 1,2, \ldots, N\}$. The adjacency matrix has elements $A_{ij}$ defined by
$$
A_{ij} = \begin{cases}
1 & \text{if } (v_i,v_j) \in \mathcal{L} \, ,\\
0 & \text{otherwise}
\end{cases}.
$$
We let $k_i$ be the degree of node $i$, given by
$$
k_i = \sum_{j \in [\mathcal{V}]} A_{ij} 
\qquad \Rightarrow \qquad
p_k = \frac1N \sum_{i\in [\mathcal{V}]} \mathbb{I}_{\{k_i = k\}} \, ,
$$
where $p_k$ is the proportion of nodes with degree $k$.

\subsection{The stochastic SIS network epidemic}

The SIS model is one of the first ever models proposed in mathematical epidemiology \cite{Ross:1916}, and remains one of the simplest descriptions of an endemic pathogen, motivating its use here. To model SIS dynamics on a network, we let $X_i(t)$ be a random variable for the state of the $i$-th node taking values in $\{S, I\}$ at time $t$ and with law evolving according to
\begin{subequations} 
\begin{align}
\mathrm{Pr}(X_i(t + \delta t) = I | X_i(t) = S) & = \tau \delta t \sum_{j \in [\mathcal{V}]} A_{ij} \mathbb{I}_{\{X_j = I\}} + o (\delta t) \, , \\
\mathrm{Pr}(X_i(t + \delta t) = S | X_i(t) = I) & = \gamma \delta t + o (\delta t), 
\end{align} \label{eqn-SIS-full}
\end{subequations}
where $\tau$ denotes the rate of transmission per link and $\gamma$ denotes the rate of recovery. These equations define a continuous-time Markov chain that we simulate using the \textit{EoN} Python package \cite{Miller:2019}, and which we seek to approximate mean behaviour using a pair-approximation (pairwise) model.

\subsection{Mean-field model}

For consistency with later notation, we let expected node-level prevalences be written like
\begin{equation*}
[A] = \mathbb{E}\left[\sum_{i \in [\mathcal{V}]} \mathbb{I}_{\{X_i = A\}}\right]\, ,
\qquad A \in \{S, I\} \, . 
\end{equation*}
The mean-field SIS model is then given by
\begin{equation}
\frac{\mathrm{d}}{\mathrm{d}t}[S] = - \frac{\beta}{N} [S][I] + \gamma [I] \, , \qquad
\frac{\mathrm{d}}{\mathrm{d}t}[I] = \frac{\beta}{N} [S][I] - \gamma [I] \, .
\label{sismf}
\end{equation}
Around the disease-free equilibrium, where $[S] \approx N$, this model has an exponential growth rate in $[I]$ of $\beta - \gamma$ and provided $\beta > \gamma$ there will be a mean-field endemic disease state
\begin{equation} \label{eqn-I-star-zerophi}
    [I]^*_{\mathrm{MF}} = N\left( 1 - \gamma/\beta \right) \, .
\end{equation}
To make the link between the mean-field model and the network model, we start by writing down unclosed equations for node-level prevalences. In \cite{Simon:2011}, an argument is given to go from the Chapman-Kolmogorov equations for the Markov chain Equations \eqref{eqn-SIS-full} to equations for the expectations of pairs
\begin{equation*}
[AB] = \mathbb{E}\left[\sum_{i,j\in [\mathcal{V}]} A_{ij} \mathbb{I}_{\{X_i = A \& X_j = B\}}\right]\, ,
\qquad A, B \in \{S, I\} \, . 
\end{equation*}
We remind that by an implicit direction, the status pair $BA$ is counted in the same way as $AB$. Hence, these unclosed equations are
\begin{equation}
\frac{\mathrm{d}}{\mathrm{d}t}[S] = - \tau [SI] + \gamma [I] \, , \qquad
\frac{\mathrm{d}}{\mathrm{d}t}[I] = \tau [SI] - \gamma [I] \, .
\label{sisunc}
\end{equation}
The simplest closure for these equations is to approximate
\begin{equation}
  [SI] \approx \frac{n}{N} [S] [I] \, ,
\label{simpleclose}
\end{equation}
where $n$ is the mean links per node. Substituting Equation \eqref{simpleclose} into \eqref{sisunc} and comparing it to Equation \eqref{sismf} implies that $\beta=n\tau$. This is the (standard) assumption that we make to use the mean-field model to approximate network dynamics \cite{House:2011}.

\subsection{Pairwise epidemic model}

We now develop a pair approximation to network SIS dynamics using notation based on \cite{Eames:2002}. The following differential equations represent the \emph{unclosed} SIS pairwise model. Each equation below describes the rate of change for expected pairs derived from all transitions related to that pair. The system remains unclosed due to the existence of triple-level quantities which are defined as expectations in the same manner as the pairs.
\begin{subequations} 
\begin{align}
\frac{\mathrm{d}}{\mathrm{d}t}[SS] &= -2\tau [SSI] + 2\gamma [SI] \, , \label{eqn-SIS-pairwise-SS} \\
\frac{\mathrm{d}}{\mathrm{d}t}[SI] &= \tau [SSI] - \tau [ISI] - \gamma [SI] -\tau[SI] +\gamma [II] \, , \label{eqn-SIS-pairwise-SI} \\
\frac{\mathrm{d}}{\mathrm{d}t}[II] &= 2\tau [ISI] + 2\tau[SI] - 2\gamma[II] \, .\label{eqn-SIS-pairwise-II}
\end{align} \label{eqn-SIS-pairwise}
\end{subequations}
A moment-closure method will now be presented to approximate the triples that are given by a function of the lower-level node- and pair-level quantities. This approximation here follows the motif expansion approach of \cite{House:2009}. Now when modifying the model to be more complex, several parameters will be introduced including a parameter controlling the network heterogeneity,
\begin{equation}
    \kappa := \frac{\sum_k k (k-1)p_k }{\left(\sum_l l p_l \right)^2} 
    = 1 + \frac{\mathrm{var}(k)}{\mathrm{mean}(k)^2} - \frac{1}{\mathrm{mean}(k)}, 
    \label{kappadef}
\end{equation}
where $\mathrm{mean}(k)$ and $\mathrm{var}(k)$ are the mean and variance of the 
network degree distribution respectively. In this way, $\kappa$ encodes variability
in the network degree distribution, taking the value of $1$ for a Poisson distribution,
and larger values for larger variance.
For a network without triangle motifs, we propose the closure estimation by the following standard approximation:
\begin{equation}
    [ABC] \approx \kappa\frac{[AB][BC]}{[B]}. \label{kirkunc}
\end{equation}
We will also present the results on three distinct networks. In previous works, the expression of $\kappa$ has been calculated for the Poisson Erd\H{o}s-R\'{e}nyi network \cite{Rand:1999} and a network with a fixed value of neighbours per node, usually called $n$-regular, shown in \cite{Keeling:1997} with further explanation \cite{Keeling:1999}. Now, when there are non-negligible probabilities of higher-order motifs, particularly triangles, existing in the network, the clustered `Kirkwood' approximation \cite{Kirkwood:1942} proposed for epidemics by Morris \cite{Morris:1997} is necessary for a more accurate approximation of triple prevalence
\begin{equation}
    [ABC] \approx \kappa\frac{[AB][BC]}{[B]}(1-\phi) + \frac{\kappa N}{n}\frac{[AB][BC][AC]}{[A][B][C]}\phi \, ,
\end{equation}
where we use $\phi$ for the clustering coefficient that is equal to the triangle appearance probability. 

Note that these moment-closure approximations describe triples in terms of pairs and singletons. Therefore, to close System \eqref{eqn-SIS-pairwise}, we need to describe the dynamics of the singletons $[S]$ and $[I]$. These are given by the System \eqref{sisunc} above, which need to be considered alongside the System \eqref{eqn-SIS-pairwise}.

Once System \eqref{eqn-SIS-pairwise} has been closed, we can consider the existence of a steady-state solution for the three pair variables, $[SS]$, $[SI]$ and $[II]$. For homogeneous pairwise equations that seek to approximate epidemics on $n$-regular networks, the solutions at the equilibrium are derived in \cite{Kiss:2017}.  In general, it is possible to get an expression in terms of $[I]^*$ (or $[S]^*$) for a particular network when considering zero clustering coefficient. The reason for this tractability is that the pair-level quantities must sum up to the number of links in a network, which is a conserved quantity. For a network with degree $n$ on average for each node, we use counting conventions in which the number of links $[-]$ will equal $Nn$. Using this and assuming zero clustering coefficient ($\phi=0$), we can calculate the endemic prevalence for a more general degree distribution than in \cite{Kiss:2017} as
\begin{align} \label{eqn-I-star-zerophi}
    [I]^*_{\phi=0} &= \frac{ -2\gamma N(\kappa-1)-\kappa\tau(N+[-]) \pm \kappa \sqrt{4\gamma N\tau(\kappa[-]-[-]+N) + \tau^2(N-[-])^2}}{2\big(\gamma(\kappa-1)^2-\kappa\tau \big)}\\
    &=\frac{-2\gamma N(\kappa-1)-\kappa N\tau(n+1) \pm \kappa N\sqrt{ 4\gamma\tau(\kappa n - n+1)+\tau^2(n-1)^2 }}{2\big(\gamma(\kappa-1)^2-\kappa\tau \big) }. \nonumber
\end{align}
The derivation of this result is given in Appendix~\ref{app:endemic}. We note that other work considering more general degree distributions than $n$-regular typically used variables like $[A_kB_l]$ for the number of pairs with a node of degree $k$ with disease state $A$ and node of degree $l$ with state $B$ \cite{Eames:2002, House:2011, Rattana:2014}. In contrast, our approach uses the variable $\kappa$ as defined in Equation \eqref{kappadef} to capture heterogeneity in the network degree distribution, leading to the closed-form solution in Equation \eqref{eqn-I-star-zerophi}.

To determine whether the epidemic will reach either the endemic equilibrium or extinction, we must consider the stability of the solutions. To investigate the stability of the endemic equilibrium, we consider the stability of the disease-free state. Following \cite{Hethcote:1987}, if the disease-free state is stable, there is no endemic equilibrium, and if the disease-free state is unstable, the endemic equilibrium exists and is stable. To investigate the stability of the disease-free state, we calculated the Jacobian of the linearised model about it; this is analysed in Appendix~\ref{app:jacobian}, and the results show instability of the disease-free equilibrium if there is a biologically reasonable solution to Equation \eqref{eqn-I-star-zerophi}. 

We now explicitly consider the influence of different network structures on the long-term expected number of infected individuals, given by the endemic equilibrium Equation \eqref{eqn-I-star-zerophi}. Since the disease only spreads if the basic reproduction number, $\mathcal{R}_0$, for a network is bigger than 1, below which the endemic equilibrium is unstable and the disease-free equilibrium is stable, the expected number of infected individuals is written as a piecewise function. Here we calculate the epidemic on three different networks: Erd\H{o}s-R\'enyi, $n$-regular and a network in which the degree of node is geometrically distributed (GDD), obtained by substituting specific values into the general \eqref{eqn-I-star-zerophi}:

\begin{align}\label{eqn-Istar-ER}
[I]^*_{\text{ER}} & = \begin{cases}
    \frac{N}{2} \Big(n+1 - \sqrt{ \frac{4\gamma}{\tau}+(n-1)^2 }\Big) & \text{if } n\tau>\gamma ,\\
    0 & \text{otherwise.}
\end{cases}\\
 \label{eqn-Istar-nreg}
[I]^*_{n\text{-reg}} & = \begin{cases}
    Nn\frac{\gamma-\tau(n-1)}{\gamma -n\tau(n-1)} & \text{if } (n-1)\tau>\gamma ,\\
    0 & \text{otherwise.}
\end{cases}\\
\label{eqn-Istar-geom}
[I]^*_{\text{GDD}} & = \begin{cases}
    Nn \Big( \frac{-\gamma(n-2) -\tau(n^2-1) + (n-1)\sqrt{4\gamma\tau(n-1)+\tau^2(n-1)^2}}{\gamma(n-2)^2-2n\tau(n-1)} \Big) & \text{if } 2(n-1)\tau>\gamma ,\\
    0 & \text{otherwise.}
\end{cases}
\end{align}

\subsection{Continuous $n$} \label{sect:cont-n}
For the Erd\H{o}s-R\'enyi and GDD networks, solutions exist for continuous values of $n$. However, we cannot have a continuous $n$-regular since such networks are not possible to construct. Instead, let us define $n = \floor{n} + \varepsilon$ and consider a \textit{minimum-variance} degree distribution with mean degree $n$,
$$
p_{\floor{n}} = 1 - \varepsilon \, , \qquad
p_{\ceil{n}} = \varepsilon \, .
$$
Here it turns out that
\begin{equation}
    \kappa = \frac{n-1}{n} + \frac{\varepsilon (1-\varepsilon)}{n^2} \, .
\label{kapeps}
\end{equation}
Therefore, to enable the interpretation of continuous values of $n$ for all networks considered, we might want to use the second term on the right-hand side as a correction to the $n$-regular results between integer values -- in practice we will interpolate between integer values using the expression in Equation \eqref{eqn-Istar-nreg} and view Equation \eqref{kapeps} as bounding the error between this and a minimum-variance degree distribution.

\subsection{Optimal contact calculation}

A network explicitly represents the connections between individuals. An individual with more connections is likely to gain increased benefits from interactions within the community, but also more likely to be infected by their neighbours or act as a superspreader. Those infected individuals probably lose a chance to work or collaborate with the community at their fullest potential, adding a cost to infection, aside from morbidity or mortality risks. Our aim in this section is to create a function describing the benefit of network interactions, with two assumptions as follows. Firstly, a network with no or few connections should give negligible benefit and infection opportunities. Secondly, the total benefit gained from interactions cannot increase without bound, instead approaching a benefit saturation point. In this benefit function, we also assume that individuals do not have more global knowledge of infection prevalence including whether or not neighbouring nodes are infected. 

Below is one of the possible functions defining a biologically reasonable (as per the criteria above) benefit of interaction function, which we investigate in the remainder of this study. This is the Hill function
\begin{equation} \label{eqn:benefit-func}
{B}(n) = \frac{bn^\alpha}{n^\alpha+n_{50}^\alpha},
\end{equation}
where $b$ denotes the maximal level of benefit, $n_{50}$ is the number of contacts resulting in half value of benefit $b$, and $\alpha$ is an exponent controlling curve steepness. In this definition, this function does not explicitly depend on the network being considered. There are two potential approaches to evaluate this benefit function on a given network. 

Firstly, where we have a population with a proportion $p_k$ with $k$ contacts, we can evaluate the benefits at an individual level by calculating the average per-person benefit, which in this context we call the mean of benefit function of interaction $\mathscr{B}_\text{MOF}$, i.e.
\begin{equation} \label{eqn:benefit-MOF}
\mathscr{B}_\text{MOF} = \langle {B}(k) \rangle = \sum_{k=0}^{\infty} p_k {B}(k).
\end{equation}
Secondly, if we evaluate the benefits at a population level rather than an individual level, we can calculate the benefit associated with the average connectivity in the population, which we refer to as the benefit function of mean links $\mathscr{B}_\text{FOM}$, i.e.
\begin{equation} \label{eqn:benefit-FOM}
\mathscr{B}_\text{FOM} = B(\langle k \rangle) \, ,
\quad \text{where}
\ \langle k \rangle = \sum_{k=0}^{\infty} p_k k \, .
\end{equation}
Such a function of mean might be most appropriate to capture effects where benefits are obtained from the overall structure of the network -- e.g. path lengths between individuals are typically reduced as $\langle k \rangle$ increases.
When the three networks considered have the same mean degree, the mean of the benefit function will have different values for each network, whereas the benefit function of mean links will have the same value among the three networks. In this study, the cost of infection can be realised as a negative benefit to an individual. Therefore, we define the total cost-benefit function
\begin{equation} \label{eqn:total-cost}
    \mathscr{C} = \mathscr{B} - [I]^* \, .
\end{equation}
% TH: TO DO: Here and elsewhere should probably change to [I]^*/N
% and be more clear we consider PER CAPITA costs and benefits.
Note that any cost constant associated with infection will be equivalent to an overall rescaling of $\mathscr{C}$ meaning that we can use the unscaled prevalence in Equation \eqref{eqn:total-cost} without loss of generality.

\section{Numerical results}

\subsection{Mean-field and SIS pairwise model comparison}

\begin{figure}[!ht]
		\centering
		\includegraphics[width=0.8\linewidth]{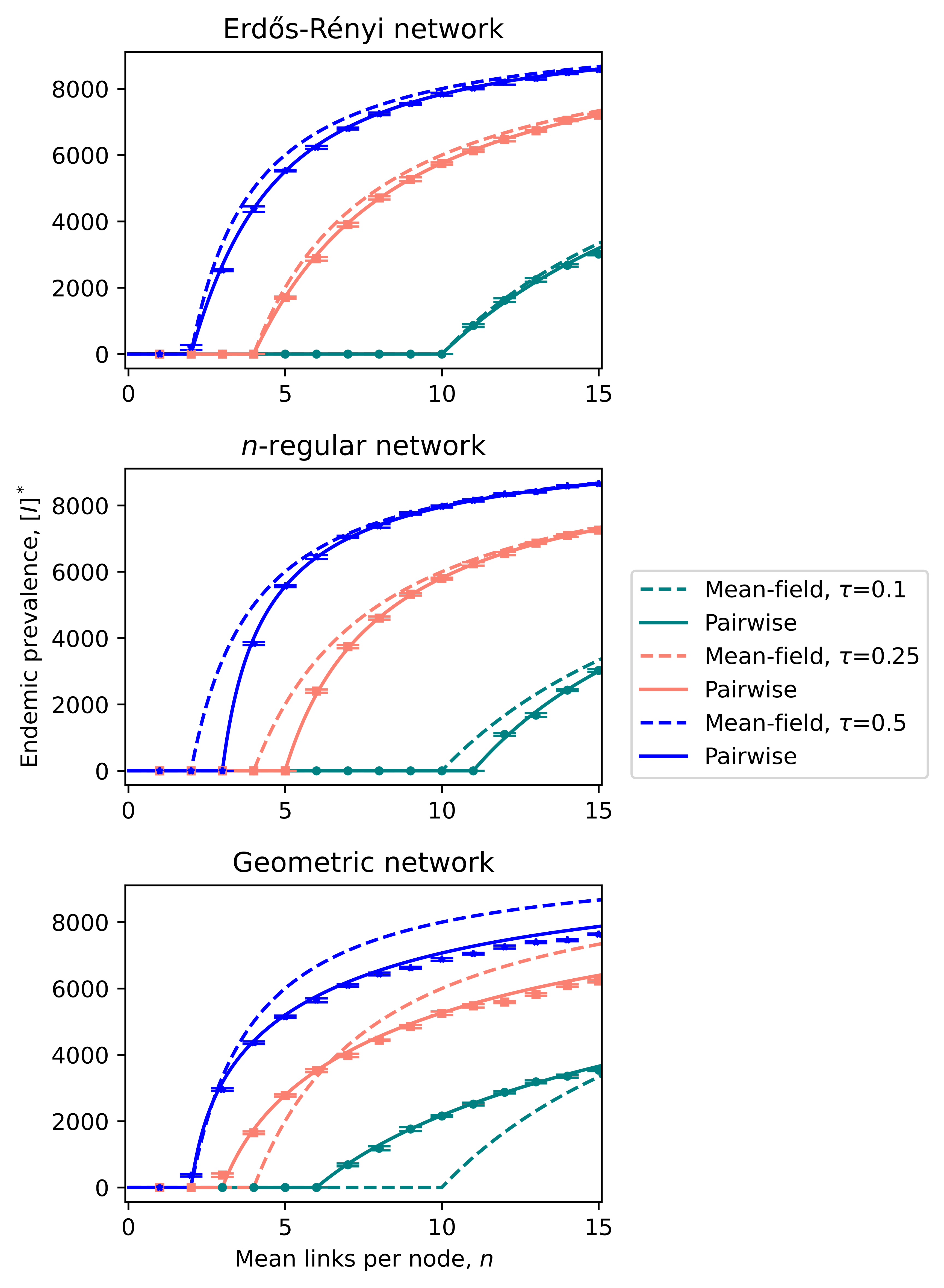}\\
		\caption{The number of infected population with mean links per node $n$ and transmission rate $\tau$ variation. Assuming the recovery rate $\gamma=1$ and the number of population $N=10000$. The dashes show the SIS mean-field solution of endemic prevalence $[I]^*$ while the solid lines lying underneath are produced from an analytical result of the SIS pairwise model on Erd\H{o}s-R\'enyi {(Top)}, $n$-regular {(Middle)}, and GDD network {(Bottom)}. Boxplots generated from thinned stochastic results when running epidemics on a network at specified mean links and $\tau$ values within time duration up to $T=1000$. } 
		\label{fig:boxplots-istar-networks}
	\end{figure}

Here, we compare the endemic equilibrium solutions to the mean-field and pairwise SIS models with the long-term quasi-steady states observed from stochastic simulations. To reduce uncertainty in the results due to multiple varying parameters, in this study we investigate the influence of changing the rate of infection $\tau$ and assume the recovery rate $\gamma=1$ since different values of this rate be absorbed into a rescaling of time. Figure \ref{fig:boxplots-istar-networks} illustrates a comparison between the pairwise epidemic model and the mean-field model. It is roughly seen that the curves for both mean-field and pairwise models tend to merge for larger mean links per node $n$ when introducing disease spreading on Erd\H{o}s-R\'enyi and $n$-regular networks. In other words, $\Delta[I]^*:=[I]^*_{g}-[I]^*_{\mathrm{MF}} \rightarrow 0$ as $n$ increases, where $g$ denotes the network type. Note that the geometric degree distribution (GDD) network takes a higher number of mean links to have a relatively close solution to the mean-field. We also notice that since the pairwise model is dependent on the network variation as a consequence of its dependence on $\kappa$, the infection threshold condition is different for each network and might be different from that of the mean-field model. The GDD network in the Figure shows how the threshold $n$ can be either the same or different between the two models depending on the value of $\tau$. In terms of errors in the pairwise model, we see the least accuracy for the Geometric degree distribution, which we would expect since network degree distribution heterogeneity coupled with SIS dynamics is complex and often counter-intuitive \cite{Berger:2005}, and we have made strong simplifying assumptions for analytical tractability, however at the level of heterogeneity associated with the Geometric distribution the level of error is acceptably small for our purposes.

The plots also verify that our pairwise model solutions derived in Equation \eqref{eqn-Istar-ER}$-$\eqref{eqn-Istar-geom} accurately approximate the full stochastic model since the boxplots align well together with the pairwise solution lines. These boxplots are generated by simulating the stochastic epidemic and randomly sampling 50 values of the number of infected individuals ($[I]^*$) in the long-term dynamics.

\begin{figure}[!ht]
		\centering
		\includegraphics[width=0.6\linewidth]{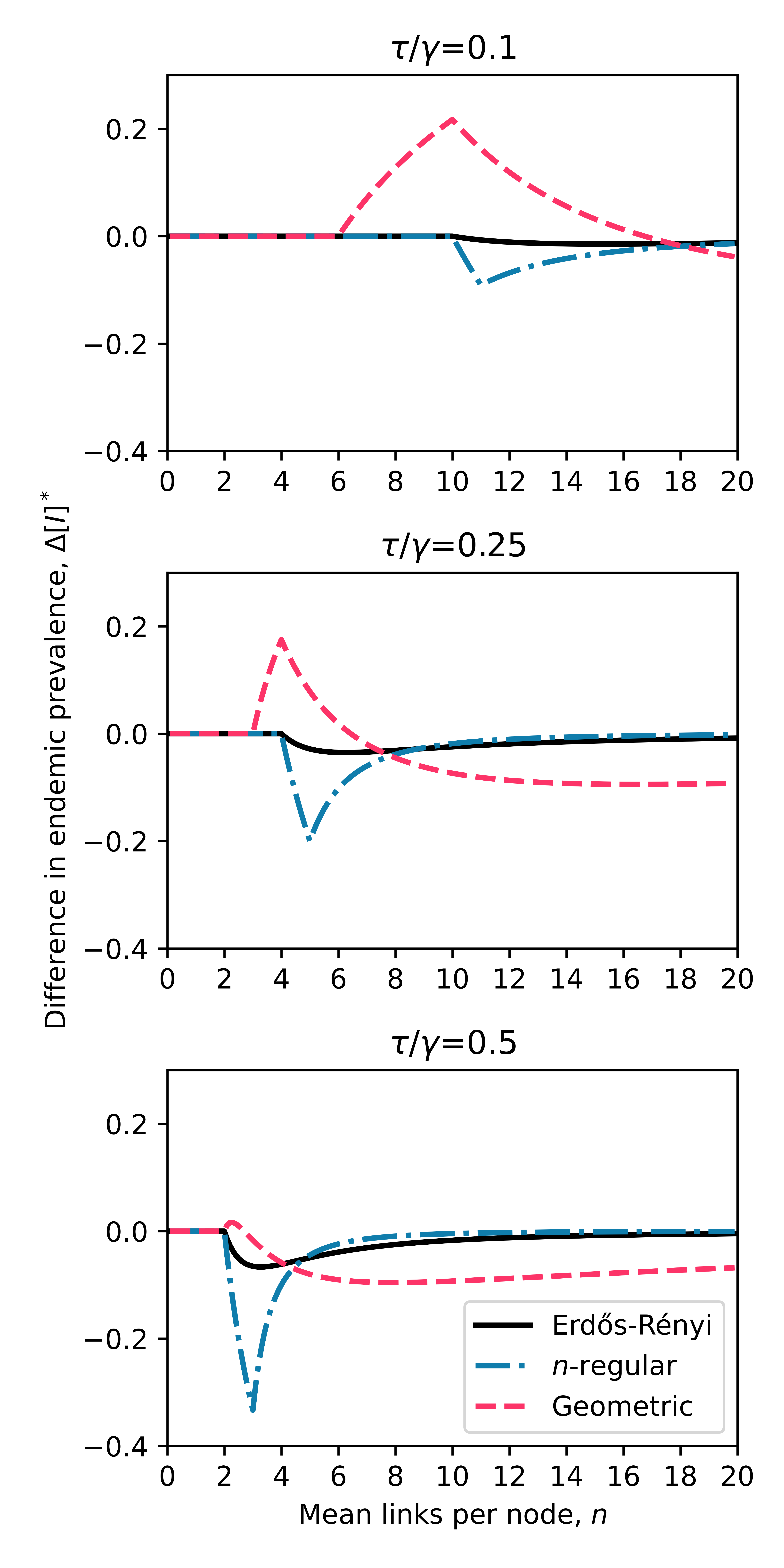}\\
		\caption{The difference in endemic prevalence per individual between the mean-field and pairwise models on each network with different $\tau$.} 
		\label{fig:istar-diff-infs}
	\end{figure}

We illustrate the difference between the two solutions more explicitly, as mentioned $\Delta[I]^*$, in Figure \ref{fig:istar-diff-infs}. The different methods yield distinct infection thresholds, with sharp rises and drops displayed in regions between two critical points. The magnitude of these changes is solely caused by the rate of infection.

\begin{figure}[!ht]
		\centering
		\includegraphics[width=0.6\linewidth]{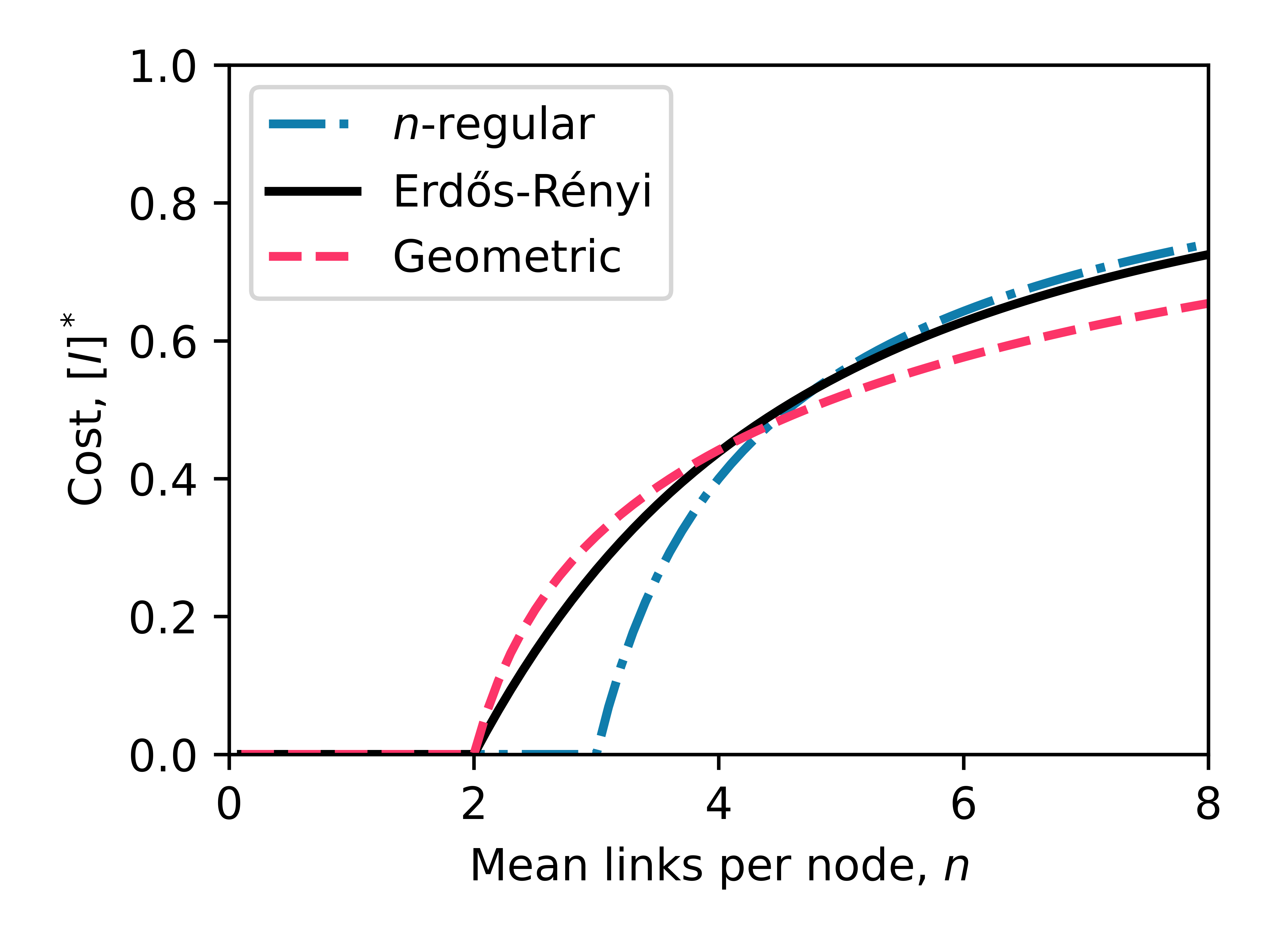}\\
		\caption{The cost of infection for each network which is proportional to the solution of the pairwise model. Here we show the curves with the transmission rate $\tau=0.50$ and recovery rate $\gamma=1$.}  
		\label{fig:cost-istar-allnets}
	\end{figure}

Comparing network heterogeneity is a key focus of this study. Regarding Figure \ref{fig:boxplots-istar-networks}, we observe the impact of different values of transmission rate on the spread of infection. To focus on the impact of network structure, measured through varying the mean link value $n$, we now focus on a single value. In Figure \ref{fig:cost-istar-allnets}, we plot each curve from Figure \ref{fig:boxplots-istar-networks} for $\tau=0.50$,  defining $[I]^*$ as the cost of infection. This cost is defined as the cost per individual, calculated as the proportion of the population who are infected in the steady state of the pairwise epidemic model. We obtain that in the $n$-regular graph if each individual has fewer than four contacts, the cost is lower than the Erd\H{o}s-R\'{e}nyi and GDD networks. Meanwhile, if all have more connections, the cost is higher than the other two networks. On the contrary, the GDD network gives us an efficient network to pay less cost per contact for a higher number of contacts, but with higher cost at lower connectivity. This cost of infection will be used together with the benefit of interaction functions to analyse the total cost-benefit in upcoming sections.

\subsection{Introducing the benefit of interaction functions}
\begin{figure}[!ht]
		\centering
        \includegraphics[width=0.6\linewidth]{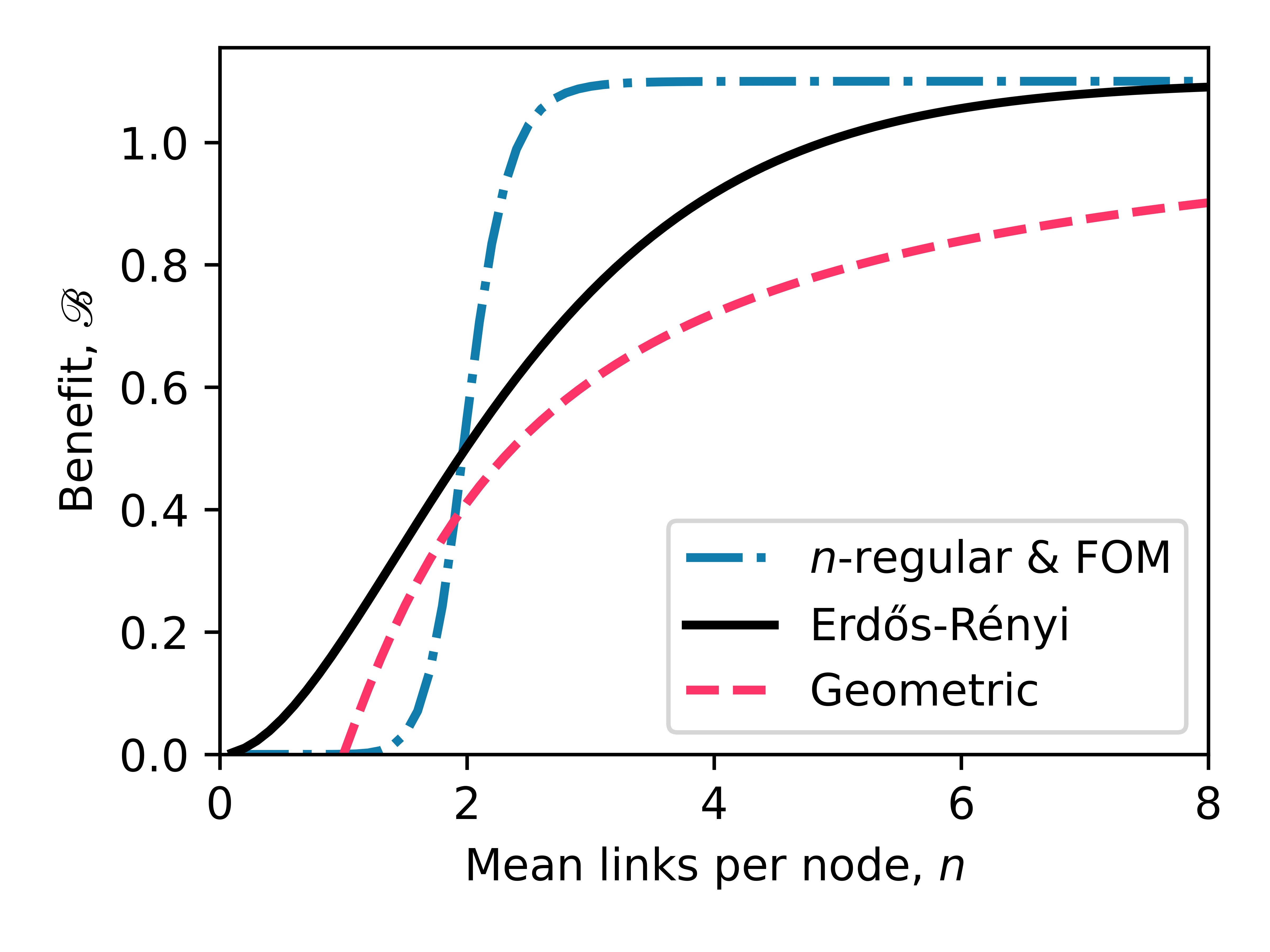}
		\caption{The benefit function in different definitions with values: benefit saturation point $b=1.10$, steep exponent $\alpha=12$, and the mean link number giving half benefit $n_{50}=2$.  As explained the benefit function of mean links (FOM) and the mean of benefit function for $n$-regular network give the same curves.}  
		\label{fig:benefit-fom-mof}
	\end{figure}

The benefit of interaction functions described by Equations \eqref{eqn:benefit-MOF} and \eqref{eqn:benefit-FOM} is illustrated in Figure \ref{fig:benefit-fom-mof}. The benefit of interaction function of mean links $\mathscr{B}_{FOM}$ depends only on the mean degree and is independent of the network heterogeneity. Therefore, for all networks, we obtain the same sigmoidal curve, and the benefit between individuals will be low when mean links $n$ is small and increases towards the maximum value $b$ when $n$ grows.

For the $n$-regular network, we have the same degree number for all nodes, and the mean of benefit function of interaction for this particular network is equivalent to the benefit function of mean links. Considering the mean of benefit function of interaction function $\mathscr{B}_{MOF}$, the Erd\H{o}s-R\'{e}nyi network gives a higher benefit followed by the GDD and $n$-regular network respectively at small values of $n$. When the networks have more connections than $n_{50}$, the $n$-regular network results in providing the highest benefit, followed by Erd\H{o}s-R\'{e}nyi and GDD networks. 

The benefit value will eventually asymptote to $b$ as the mean links $n$ becomes large. From Figure \ref{fig:benefit-fom-mof}, with a designated set of parameters, we see that the $n$-regular network with $n=3$ is already very close to this maximum benefit. For Erd\H{o}s-R\'{e}nyi network, we become very close to the maximum value $b$ around $n=8$. The GDD network however demonstrates much slower convergence to the saturation point compared to the two networks. This is because of the saturation within the benefit function, with the benefits provided by a small number of highly connected nodes not outweighing the lower benefits provided to the sparsely connected nodes. Therefore, for the majority of possible mean link values, the GDD network gives the smallest benefit value. 

\subsection{Determining an optimal contact number and its extreme values}

\begin{figure}[!ht]
		\centering
		\includegraphics[width=0.6\linewidth]{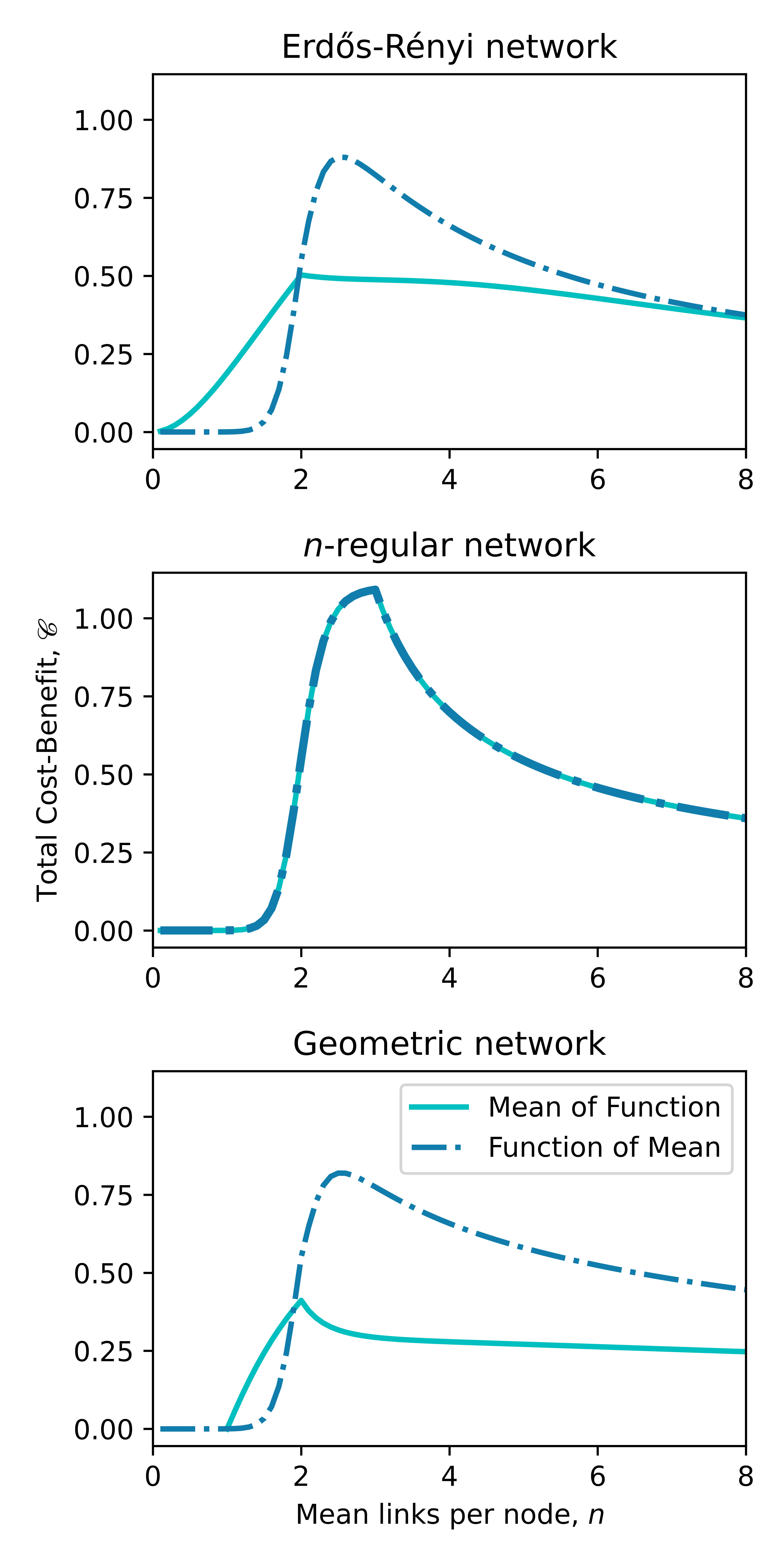}
		\caption{Total cost-benefit defined in Equation (\ref{eqn:total-cost}) using
        the pairwise approximation for endemic prevalence, for each network in comparison of the benefit of interaction function of mean links $\mathscr{B}_{FOM}$ and the mean of benefit of interaction function $\mathscr{B}_{MOF}$ with the disease transmission rate $\tau=0.50$ and recovery rate $\gamma=1$.} 
		\label{fig:total-cost-func}
	\end{figure}

\begin{figure}[!ht]
\centering
\includegraphics[width=\linewidth]{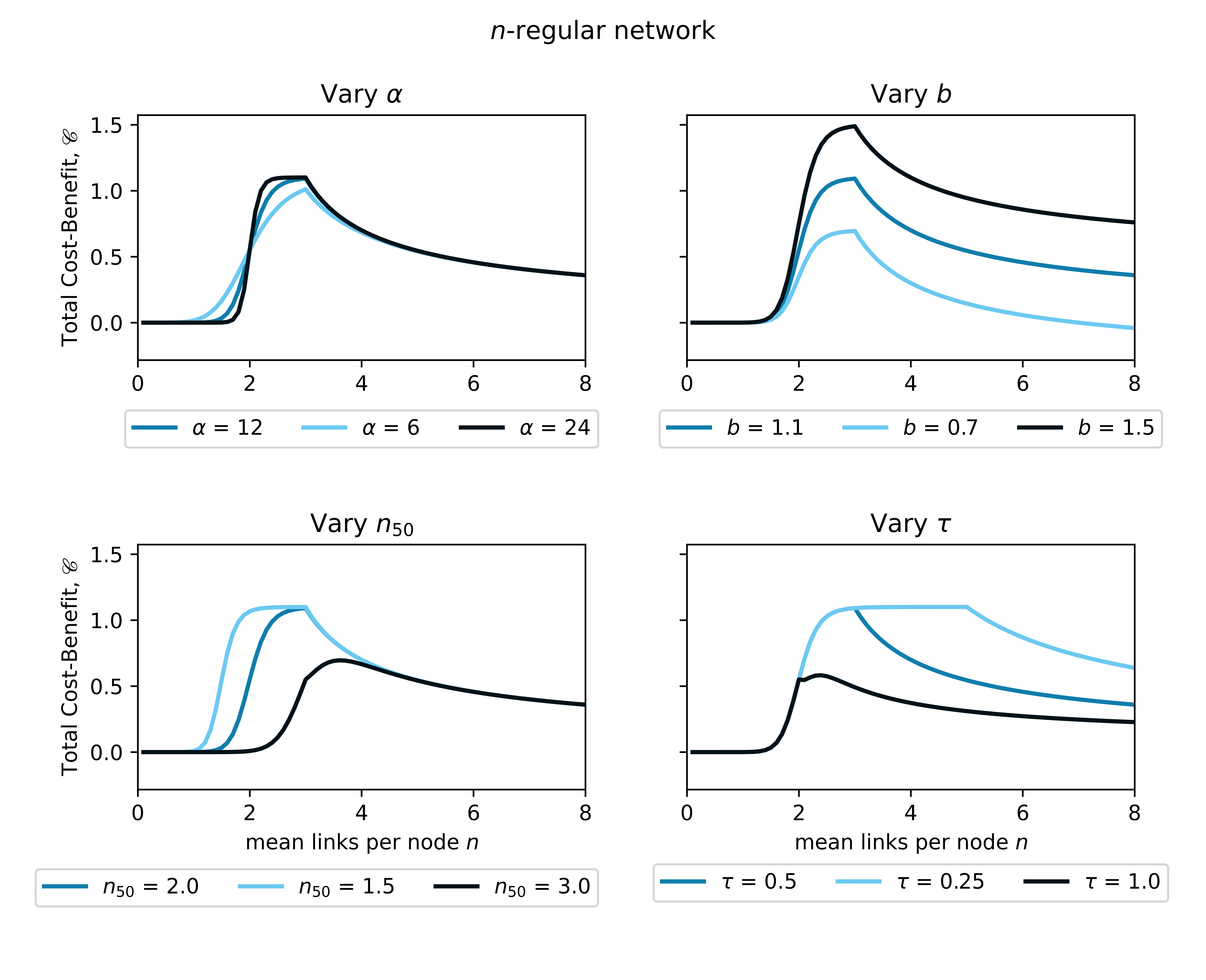}
\caption{Variability of total cost-benefit $\mathscr{C}$ for $n$-regular network and parameter values as labelled in the figure. Both benefit functions are identical, as shown in Figure \ref{fig:total-cost-func}.} 
\label{fig:ranges-n-reg}
\end{figure}

\begin{figure}[!ht]
\centering
\includegraphics[width=.82\linewidth]{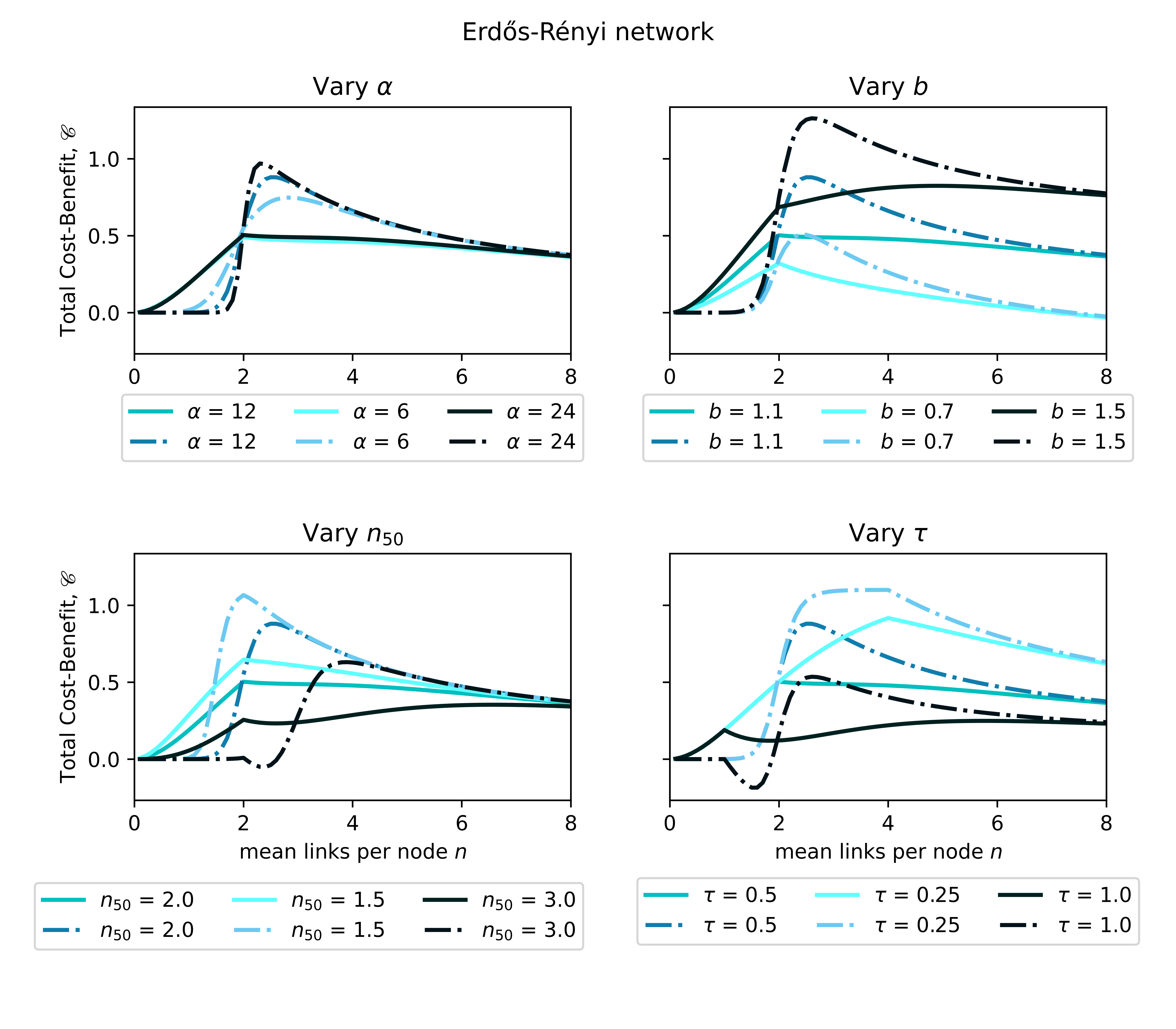}
\includegraphics[width=.82\linewidth]{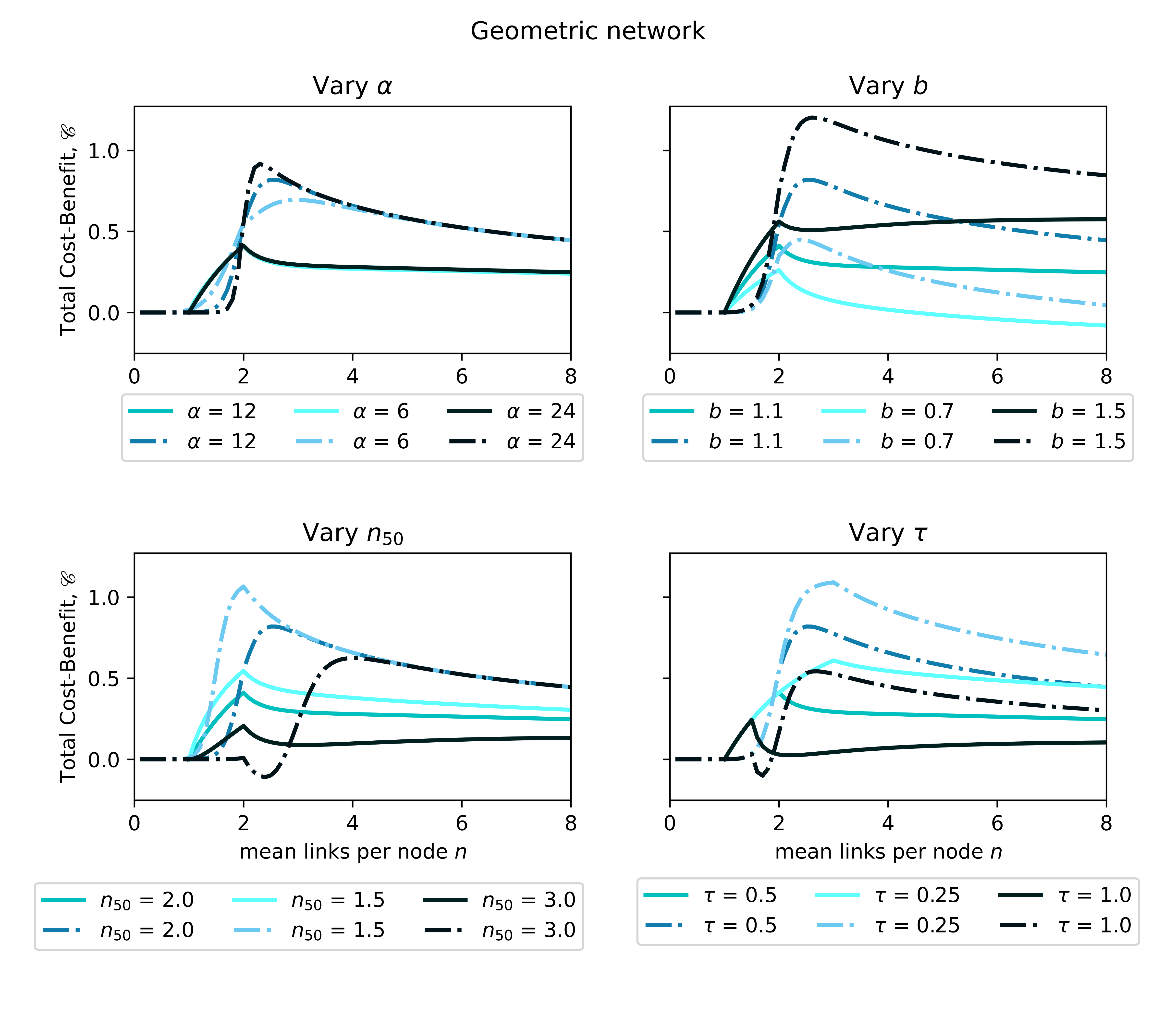}
\caption{Variability of total cost-benefit $\mathscr{C}$ for Erd\H{o}s-R\'enyi (top) and GDD (bottom) networks and parameter values as labelled in the figure.
Mean of function are shown as solid lines and function of mean as dashed-dotted lines.} 
\label{fig:ranges-ER-geom}
\end{figure}

Here, we determine the total cost-benefit, which is shown in Equation \eqref{eqn:total-cost}. Figure \ref{fig:total-cost-func} presents the number of mean contacts that maximises the cost-benefit function for individuals (restricted to mean links above the epidemic threshold). 
Figures \ref{fig:ranges-n-reg} and \ref{fig:ranges-ER-geom} show how the total cost-benefit function curves change at different parameter values, which is as would be expected from inspection of the analytic results obtained: higher $\alpha$ makes the benefit function steeper; $b$ determines maximum benefit; $n_{50}$ determines the point of inflection of the benefit function; and $\tau$ increases disease burden and hence costs.

Since the benefit of interaction for both functions is the same for $n$-regular network, the total cost is consequently similar to each other. Although the value $n$ for this network must be an integer, hence its benefit $\mathscr{B}_{MOF}$ is normally non-continuous, as suggested in Section \ref{sect:cont-n}, we can assume the benefit to be continuous. For the illustrated set of parameters, the $n$-regular network with $n=3$ is not only the optimal value for that network but also has a higher total benefit than the other two networks considered, with GDD having the lowest total benefit. 

Regardless of the magnitude of the total cost-benefit, the benefit of interaction function of mean links $\mathscr{B}_{FOM}$ provides no significant changes in optimal contact numbers for all three networks. Considering the mean of benefit of interaction function, in Erd\H{o}s-R\'{e}nyi and GDD networks, the maximum benefit $\mathscr{B}_{MOF}$ is gained for lower mean links than that of FOM. 

The benefit functions explored here yield a local maximum in the cost-benefit function since the benefit function reaches a plateau for lower values of $n$ than the endemic prevalence. Benefit functions without a local maximum also exist, where the optimal mean number of links will be a fully connected population.    

\section{Discussion}

\subsection{Conclusions}

We have developed a simple model of SIS network epidemics and obtained a closed-form solution for its endemic equilibrium. This solution is not only an efficient way to include the network heterogeneity but also should in future allow for consideration of the existence of non-zero clustering coefficient. 

This approach allows us to consider optimal trade-offs between infection and interaction as a function of network parameters such as the mean degree $n$. These network parameters are not straightforward to optimise using e.g.\ Monte Carlo methodology, but optima can be straightforwardly read off graphs such as those in Figure \ref{fig:total-cost-func}.

While the system we have considered is a significant simplification of real-world infection-interaction trade-offs, it is one where we can understand all of the model components and the contribution they make to the optimal solution of the trade-off. In this way we hope it provides a contribution to the project of understanding the complex interactions between disease and behaviour.

\subsection{Limitations and directions for future work}

In this analysis, we have defined an arbitrary benefit function and explored the relationship between the benefit granted and the cost of increased infection due to increased network connectivity. The benefit function was derived using logical arguments about the nature of social interactions. However, the exact structure is not underpinned by any data, and in reality such benefit functions are likely to be more complex than that considered. Despite this, the methodological developments allowing the analysis of the cost-benefit function could readily be applied to any benefit functions, provided they can be described as a function of the degree distribution of a network. Future work looking at potential benefit functions, particularly in the presence factors such as individual agents acting against collective good, would therefore be necessary to translate this work to more applied settings. 

We have only considered SIS epidemic models in this analysis since we were able to obtain analytically tractable solutions to the pair-approximation methods. The cost-benefit trade-off can be performed for any epidemic model, such as SIRS, subject to being able to efficiently calculate the endemic prevalence as a function of the mean degree. 

\section*{Acknowledgements}
The authors would like to thank Edward Hill for the personal discussion of the preliminary results. We also thank the peer reviewers whose constructive comments improved this manuscript for result consistency.

\section*{Funding}
B.T. is financially supported by the Development and Promotion of Science and Technology Talents Project (DPST) jointly provided by the Institute for the Promotion of Teaching Science and Technology (IPST), Thailand and the Royal Thai Government. He is also supported by the European Society for Mathematical and Theoretical Biology (ESMTB). T.H. is supported by the Wellcome Trust (award 227438/Z/23/Z) and UK Research and Innovation.

\section*{Conflict of Interests}
The authors declare no conflict of interests.

\section*{Code availability}
The code needed to reproduce the Figures in this article is available at
https://github.com/thatchai-b/article-trade-off-infs-social-interacts .

\appendix

\section{Endemic infection}

\label{app:endemic}

According to the conserved property, one equation of the System \eqref{eqn-SIS-pairwise} can be negligible. Assuming non-zero infection $[SI]^*\ne0$, we know that $[SI]^* = \frac{\gamma}{\tau}$ and we also obtain the pairwise numbers at the equilibrium as the following
\begin{equation}
	\begin{aligned}
		\gamma[S] &= \kappa\tau[SS]\bigg[1-\phi+\frac{N}{n}\frac{[SI]}{[S][I]}\phi\bigg] \nonumber \\
		\Rightarrow \quad [SS]^* &= \frac{\gamma n(N-[I]^*)^2}{\kappa n \tau(N-[I]^*)(1-\phi)+\gamma\kappa N\phi},
	\end{aligned}
\end{equation}
together with
 \begin{equation*}
	\begin{aligned}
		\gamma[II] &= \tau [ISI] + \tau[SI]\\
		\Rightarrow \quad [II]^* &= \frac{ n\tau(N-[I]^*)[I]^*+\gamma\kappa n([I]^*)^2(1-\phi) }{ n\tau(N-[I]^*) - \gamma\kappa N\phi }. \\
		\end{aligned}
	\end{equation*}
Assuming a zero clustering coefficient, we have the numbers 
\begin{subequations}
	\begin{align}
		[SS]^*_{\phi=0} &= \frac{\gamma}{\kappa\tau} (N-[I]^*) \label{eqn-SIS-pairwise-SS-zerophi} \\
		[SI]^*_{\phi=0} &= \frac{\gamma}{\tau} [I]^* \label{eqn-SIS-pairwise-SI-zerophi} \\
		[II]^*_{\phi=0} &= [I]^* + \frac{\gamma\kappa}{\tau}\frac{([I]^*)^2}{N-[I]^*}. \label{eqn-SIS-pairwise-II-zerophi}
	\end{align}
\end{subequations}
Substituting these pairwise solutions into $[-]=[SS]+2[SI]+[II]$ gives a quadratic polynomial that can be solved for an endemic equilibrium shown in Equation \eqref{eqn-I-star-zerophi}.

\section{Stability of the disease-free state}

\label{app:jacobian}

The stability of the disease-free equilibrium can be analysed using the Eigenvalues of the following Jacobian matrix
	\begin{equation*}
		\boldsymbol{J} = \begin{bmatrix}
			-\gamma & \tau & 0 \\
			g_{[I]} & g_{[SI]} & g_{[II]} \\
			\frac{2\kappa\tau[SI]^2}{(N-[I])^2}\big[1-\phi+\frac{N}{n}[II]\big(\frac{3[I]-2N}{[I]^3}\big)\phi\big] & \frac{4\kappa\tau[SI]}{N-[I]}\big[1-\phi+\frac{N}{n}\frac{[II]}{[I]^2}\phi\big]+2\tau & \frac{2\kappa\tau[SI]^2}{N-[I]}\big[\frac{N}{n[I]^2}\phi \big]-2\gamma 
		\end{bmatrix}    \, ,
	\end{equation*}
	where
	\begin{equation*}
		\begin{aligned}
			g_{[I]} &= \tfrac{\kappa\tau[SI]}{(N-[I])^2}([-]-3[SI]-[II])(1-\phi) + \tfrac{\kappa N\tau[SI]^2}{n[I]^3(N-[I])^3}\big[2N^2[II]+[I](2N[SI]-[-]N-4N[II]) \\
			&+[I]^2(3[-]-6[SI])\big]\phi\, , \\
			g_{[SI]} &= \tfrac{\kappa\tau}{N-[I]}([-]-6[SI]-[II])(1-\phi)+\tfrac{2\kappa N\tau[SI]}{n[I]^2(N-[I])^2}([-][I]-3[I][SI]-N[II])\phi-\tau-\gamma \, ,\\
			g_{[II]} &= -\tfrac{\kappa\tau[SI]}{N-[I]}\big[1-\phi+\tfrac{N^2[SI]}{n[I]^2(N-[I])}\phi\big]+\gamma \, .
		\end{aligned}
	\end{equation*} 
	For simplicity, we consider the zero-$\phi$ Jacobian matrix and will then be looking for the Eigenvalues of disease-free equilibrium 
	\begin{equation*}
		\boldsymbol{J}_{\phi=0} = \begin{bmatrix}
			-\gamma & \tau & 0 \\
			\frac{\kappa\tau[SI]}{(N-[I])^2}([-]-3[SI]-[II]) & \frac{\kappa\tau}{N-[I]}([-]-6[SI]-[II])-\tau-\gamma & -\frac{\kappa\tau[SI]}{N-[I]}+\gamma \\
			\frac{2\kappa\tau[SI]^2}{(N-[I])^2} & \frac{4\kappa\tau[SI]}{N-[I]}+2\tau & -2\gamma 
		\end{bmatrix}   \, .
	\end{equation*}	

We obtain three Eigenvalues: 
$$\lambda_1=-\gamma \, , \quad 
\lambda_{2,3}=-\frac{1}{2}\Big(3\gamma+\tau-\frac{\kappa\tau}{N}[-]\Big) \pm \frac{1}{2}\sqrt{\Big(3\gamma+\tau-\frac{\kappa\tau}{N}[-]\Big)^2-4\Big(2\gamma^2-2\frac{\gamma\kappa\tau}{N}[-]\Big)}\, .
$$
It can be verified that the latter solutions are real when
\begin{equation*} 
\Big(\gamma+\frac{\kappa\tau}{N}[-]-\tau\Big)^2+8\gamma\tau \ge 0.
\end{equation*}


\begin{thebibliography}{10}

\bibitem{WHO:2024}
{World Health Organisation}.
\newblock {COVID-19 dashboard}, 2024.

\bibitem{Hale:2021}
Thomas Hale, Noam Angrist, Rafael Goldszmidt, Beatriz Kira, Anna Petherick,
  Toby Phillips, Samuel Webster, Emily Cameron-Blake, Laura Hallas, Saptarshi
  Majumdar, and Helen Tatlow.
\newblock A global panel database of pandemic policies ({Oxford} {COVID}-19
  {Government} {Response} {Tracker}).
\newblock {\em Nature Human Behaviour}, 5(4):529--538, 2021.

\bibitem{Whittles:2016}
Lilith~K. Whittles and Xavier Didelot.
\newblock Epidemiological analysis of the {E}yam plague outbreak of 1665-1666.
\newblock {\em Proceedings of the Royal Society B: Biological Sciences},
  283(1830):20160618, 2016.

\bibitem{Herrera:2019}
James Herrera and Charles~L. Nunn.
\newblock Behavioural ecology and infectious disease: implications for
  conservation of biodiversity.
\newblock {\em Philosophical Transactions of the Royal Society B: Biological
  Sciences}, 374(1781):20180054, 2019.

\bibitem{Vedhara:2022}
Kavita Vedhara, Kieran Ayling, Ru~Jia, Lucy Fairclough, Joanne~R Morling,
  Jonathan~K Ball, Holly Knight, Holly Blake, Jessica Corner, Chris Denning,
  Kirsty Bolton, Hannah Jackson, Carol Coupland, and Patrick Tighe.
\newblock Relationship between anxiety, depression, and susceptibility to
  severe acute respiratory syndrome coronavirus 2 infection: Proof of concept.
\newblock {\em The Journal of Infectious Diseases}, 225(12):2137--2141, 2022.

\bibitem{Wu:2022}
Chao-Yi Wu, Nora Mattek, Katherine Wild, Lyndsey~M. Miller, Jeffrey~A. Kaye,
  Lisa~C. Silbert, and Hiroko~H. Dodge.
\newblock Can changes in social contact (frequency and mode) mitigate low mood
  before and during the {COVID-19} pandemic? {The I-CONECT} project.
\newblock {\em Journal of the American Geriatrics Society}, 70(3):669--676,
  2022.

\bibitem{Bolton:2024}
Kirsty~J. Bolton, Armando Mendez-Villalon, Henry Nanji, Ru~Jia, Kieran Ayling,
  Grazziela Figueredo, and Kavita Vedhara.
\newblock Monitoring university student response to social distancing policy
  during the {SARS}-{CoV}-2 pandemic using {Bluetooth}: the {RADAR} study.
\newblock {\em Mathematics in Medical and Life Sciences}, 1(1):2425096,
  December 2024.

\bibitem{Eames:2002}
Ken T.~D. Eames and Matt~J. Keeling.
\newblock Modeling dynamic and network heterogeneities in the spread of
  sexually transmitted diseases.
\newblock {\em Proceedings of the National Academy of Sciences},
  99(20):13330--13335, 2002.

\bibitem{Hagenaars:2004}
T.J. Hagenaars, C.A. Donnelly, and N.M. Ferguson.
\newblock Spatial heterogeneity and the persistence of infectious diseases.
\newblock {\em Journal of Theoretical Biology}, 229(3):349--359, 2004.

\bibitem{Keeling:2005}
Matt~J Keeling and Ken~TD Eames.
\newblock Networks and epidemic models.
\newblock {\em Journal of The Royal Society Interface}, 2(4):295–307, 2005.

\bibitem{Kiss:2017}
István~Z. Kiss, Joel~C. Miller, and Péter~L. Simon.
\newblock {\em Mathematics of Epidemics on Networks: From Exact to Approximate
  Models}, volume~46 of {\em Interdisciplinary Applied Mathematics}.
\newblock Springer International Publishing, Cham, Switzerland, 2017.

\bibitem{Pastor-Satorras:2001}
Romualdo Pastor-Satorras and Alessandro Vespignani.
\newblock Epidemic spreading in scale-free networks.
\newblock {\em Physical Review Letters}, 86(14):3200--3203, 2001.

\bibitem{Ferreira:2012}
Silvio~C Ferreira, Claudio Castellano, and Romualdo Pastor-Satorras.
\newblock Epidemic thresholds of the susceptible-infected-susceptible model on
  networks: A comparison of numerical and theoretical results.
\newblock {\em Physical Review E—Statistical, Nonlinear, and Soft Matter
  Physics}, 86(4):041125, 2012.

\bibitem{Lajmanovich:1976}
Ana Lajmanovich and James~A Yorke.
\newblock A deterministic model for gonorrhea in a nonhomogeneous population.
\newblock {\em Mathematical Biosciences}, 28(3-4):221--236, 1976.

\bibitem{Sharkey:2011}
Kieran~J Sharkey.
\newblock Deterministic epidemic models on contact networks: Correlations and
  unbiological terms.
\newblock {\em Theoretical population biology}, 79(4):115--129, 2011.

\bibitem{Overton:2022}
Christopher~E Overton, Robert~R Wilkinson, Adedapo Loyinmi, Joel~C Miller, and
  Kieran~J Sharkey.
\newblock Approximating quasi-stationary behaviour in network-based sis
  dynamics.
\newblock {\em Bulletin of Mathematical Biology}, 84:1--32, 2022.

\bibitem{Andersson:2000}
H{\aa}kan Andersson and Tom Britton.
\newblock Stochastic epidemics in dynamic populations: quasi-stationarity and
  extinction.
\newblock {\em Journal of mathematical biology}, 41:559--580, 2000.

\bibitem{Kryscio:1989}
Richard~J Kryscio and Claude Lef{\`e}vre.
\newblock On the extinction of the s--i--s stochastic logistic epidemic.
\newblock {\em Journal of Applied Probability}, 26(4):685--694, 1989.

\bibitem{Naasell:1996}
Ingemar N{\aa}sell.
\newblock The quasi-stationary distribution of the closed endemic sis model.
\newblock {\em Advances in Applied Probability}, 28(3):895--932, 1996.

\bibitem{Wilkinson:2013}
Robert~R Wilkinson and Kieran~J Sharkey.
\newblock An exact relationship between invasion probability and endemic
  prevalence for markovian sis dynamics on networks.
\newblock {\em PLoS One}, 8(7):e69028, 2013.

\bibitem{Lieberman:2005}
Erez Lieberman, Christoph Hauert, and Martin~A Nowak.
\newblock Evolutionary dynamics on graphs.
\newblock {\em Nature}, 433(7023):312--316, 2005.

\bibitem{Masuda:2009}
Naoki Masuda.
\newblock Directionality of contact networks suppresses selection pressure in
  evolutionary dynamics.
\newblock {\em Journal of Theoretical Biology}, 258(2):323--334, 2009.

\bibitem{Ohtsuki:2006}
Hisashi Ohtsuki, Christoph Hauert, Erez Lieberman, and Martin~A Nowak.
\newblock A simple rule for the evolution of cooperation on graphs and social
  networks.
\newblock {\em Nature}, 441(7092):502--505, 2006.

\bibitem{Broom:2010}
Mark Broom, Christophoros Hadjichrysanthou, and Jan Rycht{\'a}{\v{r}}.
\newblock Evolutionary games on graphs and the speed of the evolutionary
  process.
\newblock {\em Proceedings of the Royal Society A: Mathematical, Physical and
  Engineering Sciences}, 466(2117):1327--1346, 2010.

\bibitem{Hindersin:2015}
Laura Hindersin and Arne Traulsen.
\newblock Most undirected random graphs are amplifiers of selection for
  birth-death dynamics, but suppressors of selection for death-birth dynamics.
\newblock {\em PLoS computational biology}, 11(11):e1004437, 2015.

\bibitem{Ross:1916}
Ronald Ross.
\newblock An application of the theory of probabilities to the study of a
  priori pathometry.--part i.
\newblock {\em Proceedings of the Royal Society A}, 92(638):204--230, 1916.

\bibitem{Miller:2019}
Joel~C. Miller and Tony Ting.
\newblock \text{EoN} (epidemics on networks): a fast, flexible \text{Python}
  package for simulation, analytic approximation, and analysis of epidemics on
  networks.
\newblock {\em Journal of Open Source Software}, 4(44):1731, 2019.

\bibitem{Simon:2011}
P{\'e}ter~L. Simon, Michael Taylor, and Istvan~Z. Kiss.
\newblock Exact epidemic models on graphs using graph-automorphism driven
  lumping.
\newblock {\em Journal of Mathematical Biology}, 62(4):479--508, 2011.

\bibitem{House:2011}
Thomas House and Matt~J. Keeling.
\newblock Insights from unifying modern approximations to infections on
  networks.
\newblock {\em Journal of The Royal Society Interface}, 8(54):67--73, 2011.

\bibitem{House:2009}
Thomas House, Geoffrey Davies, Leon Danon, and Matt~J. Keeling.
\newblock A motif-based approach to network epidemics.
\newblock {\em Bulletin of Mathematical Biology}, 71(7):1693--1706, 2009.

\bibitem{Rand:1999}
David Rand.
\newblock Correlation equations and pair approximations for spatial ecologies.
\newblock {\em CWI Quarterly}, 12(3-4):329--368, 1999.

\bibitem{Keeling:1997}
M.~J. Keeling, D.~A. Rand, and A.~J. Morris.
\newblock Correlation models for childhood epidemics.
\newblock {\em Proceedings: Biological Sciences}, 264(1385):1149--1156, 1997.

\bibitem{Keeling:1999}
Matt~J. Keeling.
\newblock The effects of local spatial structure on epidemiological invasions.
\newblock {\em Proceedings: Biological Sciences}, 266(1421):859--867, 1999.

\bibitem{Kirkwood:1942}
John~G. Kirkwood and Elizabeth~Monroe Boggs.
\newblock The radial distribution function in liquids.
\newblock {\em Journal of Chemical Physics}, 10(6):394--402, 1942.

\bibitem{Morris:1997}
Andrew~John Morris.
\newblock {\em Representing spatial interactions in simple ecological models}.
\newblock PhD thesis, University of Warwick, 1997.

\bibitem{Rattana:2014}
P.~Rattana, Joel~C. Miller, and István~Z. Kiss.
\newblock Pairwise and edge-based models of epidemic dynamics on correlated
  weighted networks.
\newblock {\em Math. Model. Nat. Phenom.}, 9(2):58--81, 2014.

\bibitem{Hethcote:1987}
Herbert~W. Hethcote and James W.~Van Ark.
\newblock Epidemiological models for heterogeneous populations: Proportionate
  mixing, parameter estimation, and immunization programs.
\newblock {\em Mathematical Biosciences}, 84:85--118, 1987.

\bibitem{Berger:2005}
Noam Berger, Christian Borgs, Jennifer Chayes, and Amin Saberi.
\newblock On the spread of viruses on the internet.
\newblock In {\em Proceedings of the 16th ACM-SIAM Symposium on Discrete
  Algorithm (SODA)}, pages 301--310, 2005.

\end{thebibliography}
\end{document}